\documentclass[letter,twocolumn]{jpsj3}

\title{
Stable Existence of Phase IV inside Phase II under Pressure in Ce$_{0.8}$La$_{0.2}$B$_{6}$
}

\author{
Keisuke \textsc{Kunimori}\thanks{E-mail address: kkunimori@hiroshima-u.ac.jp}, 
Hiroshi \textsc{Tanida}, 
Takeshi \textsc{Matsumura}, 
Masafumi \textsc{Sera}, and 
Fumitoshi \textsc{Iga}
}

\inst{
Department of Quantum Matter, AdSM, Hiroshima University, Higashi-Hiroshima, Hiroshima, 739-8530
}

\abst{
We investigate the pressure effect of the electrical resistivity and magnetization of Ce$_{0.8}$La$_{0.2}$B$_{6}$. 
The situation in which phase IV stably exists inside phase II at $H=0$ T could be realized by applying a pressure above $P\sim 1.1$ GPa. 
This originates from the fact that the stability of phase II under pressure is larger than those of phases IV and III. 
The results seem to be difficult to reproduce by taking the four interactions of $\Gamma_{\text{5u}}$-type AFO, $O_{xy}$-type AFQ, $T_{xyz}$-type AFO, and AF exchange into account within a mean-field calculation framework. 
}

\kword{Ce$_{0.8}$La$_{0.2}$B$_{6}$, phase IV, multipole order, pressure effect}

\begin{document}
\maketitle


Multipole ordering has been studied extensively as one of the topics in $f$-electron systems.\cite{ref1,ref2} 
CeB$_{6}$ is one of the most intensively studied compounds showing the unusual antiferro-quadrupole (AFQ) ordering, and now its unusual properties in the ordered phases are understood to be a result of the coexistence and competition among the $O_{xy}$-type AFQ, $T_{xyz}$-type AFO, and AF exchange interactions.\cite{ref3,ref4,ref5} 
Fifteen years ago, in Ce$_{x}$La$_{1-x}$B$_{6}$ $(x \leq 0.8)$, it was discovered that phase IV appears at $H = 0$ T as a result of the lower suppression rate of $T_{\text{N}}$ than that of $T_{\text{Q}}$ and has been studied extensively.\cite{ref6,ref7,ref8,ref9} 
The strong candidate for the order parameter in phase IV is considered to be the $\Gamma_{\text{5u}}$-type antiferro-octupole (AFO) moment from the results of a theoretical investigation and resonant X-ray and neutron diffractions.\cite{ref10,ref11,ref12,ref13} 
However, Kondo \textit{et al}. pointed out that the $O_{xy}$-type FQ ordering simultaneously induced by the $\Gamma_{\text{5u}}$-type AFO ordering competes with the $O_{xy}$-type AFQ interaction in Ce$_{x}$La$_{1-x}$B$_{6}$, and discussed the difficulty in explaining the overall properties of the Ce$_{x}$La$_{1-x}$B$_{6}$ system.\cite{ref14} 
Recently, Kondo \textit{et al}. reported that phase IV in Ce$_{x}$La$_{1-x}$B$_{6}$ ($x=0.4$ and $0.5$) is stabilized by Pr and Nd doping, which is difficult to explain by the simple $\Gamma_{\text{5u}}$-type AFO with a nonmagnetic singlet ground state. 
The results indicate that the dipole moments of Pr and Nd ions contribute to the long-range order (LRO) in phase IV.\cite{ref15} 
Thus, there still remain problems to be explained for the LRO in phase IV.

In the present study, we investigate the pressure ($P$) effect of Ce$_{0.8}$La$_{0.2}$B$_{6}$. 
This sample is situated immediately after phase IV appears.\cite{ref9} 
The ground state is phase III, and the temperature region of phase IV, which exists at the higher temperature side of phase III, is very narrow. 
The transition temperature from phase III to phase IV and that from phase IV to phase I are $T_{\text{N}} \simeq 1.6$ K and $T_{\text{IV}} \simeq1.7$ K, respectively, and phase II exists above $H\simeq 0.4$ T at $T \simeq 1.7$ K. 
In this sample, the magnitudes of the $O_{xy}$-type AFQ, $T_{xyz}$-type AFO, and AF exchange interactions, and the interaction forming phase IV are nearly the same and the last is slightly larger than the former three interactions. 
We also reported that in this sample, there exists a temperature region where the direct transition from phase IV to phase II occurs at around $H \simeq 0.4$ T and $T\simeq 1.7$ K.\cite{ref9} 
Such a direct IV-II phase transition was also reported in Ce$_{0.75}$La$_{0.25}$B$_{6}$ under the uniaxial pressure $P\parallel [001]$ for $H\parallel [110]$.\cite{ref16} 
From the study of the IV-II direct phase transition, a clue to clarifying the nature of phase IV is expected to be obtained. 
We conjecture that such a IV-II direct phase transition may be realized by applying pressure as follows. 
We reported the effect of pressure on the multipole interactions in CeB$_{6}$.\cite{ref17} 
The pressure dependence of $T_{\text{Q}}$ and $T_{\text{oct}}$ (the magnitude of the $T_{xyz}$-type AFO transition temperature) is positive and that of $T_{\text{N}}$ is negative. 
Then, phase II is stabilized by applying pressure, and finally, the situation in which phase II appears at the higher temperature side of phase IV at $H = 0$ T could be realized in Ce$_{0.8}$La$_{0.2}$B$_{6}$. 
In the present study, we investigate this direct IV-II phase transition under pressure in detail to clarify the nature of phase IV.


A single crystal of Ce$_{0.8}$La$_{0.2}$B$_{6}$ was prepared by a floating zone method using four xenon lamps.\cite{ref18} 
The specific heat was measured by a conventional thermal relaxation method (Quntum Design PPMS) down to $T\simeq 0.4$ K. 
The electrical resistivity at $H=0$ T under pressure was measured down to $0.4$ K using a conventional piston cylinder-type pressure cell. 
The magnetization was measured by a conventional extraction method under the pressure applied using a conventional piston cylinder-type pressure cell up to $ 1.2$ GPa.

\begin{figure}[t]
\begin{center}
\includegraphics[width=50mm,clip]{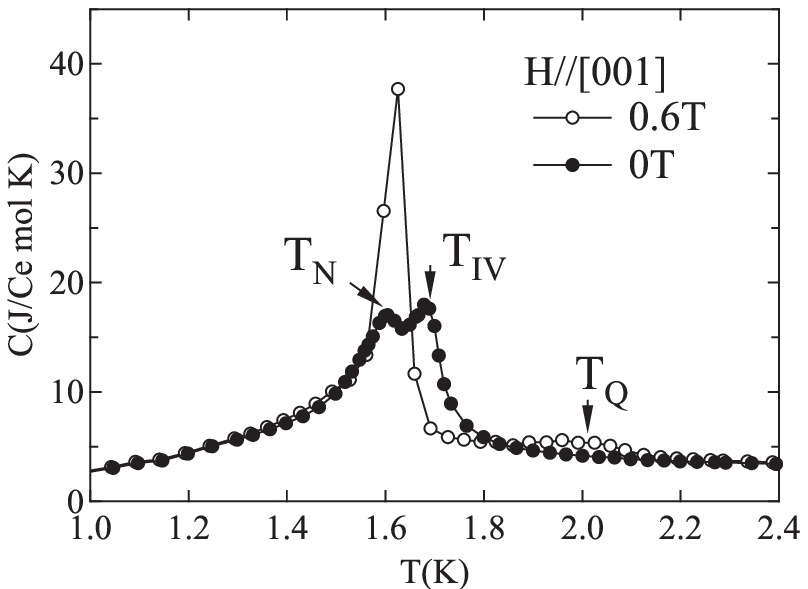}
\caption{Temperature dependence of the specific heat of Ce$_{0.8}$La$_{0.2}$B$_{6}$ at $H = 0$ and $0.6$ T for $H\parallel [001]$. }
\label{fig1}
\includegraphics[width=70mm,clip]{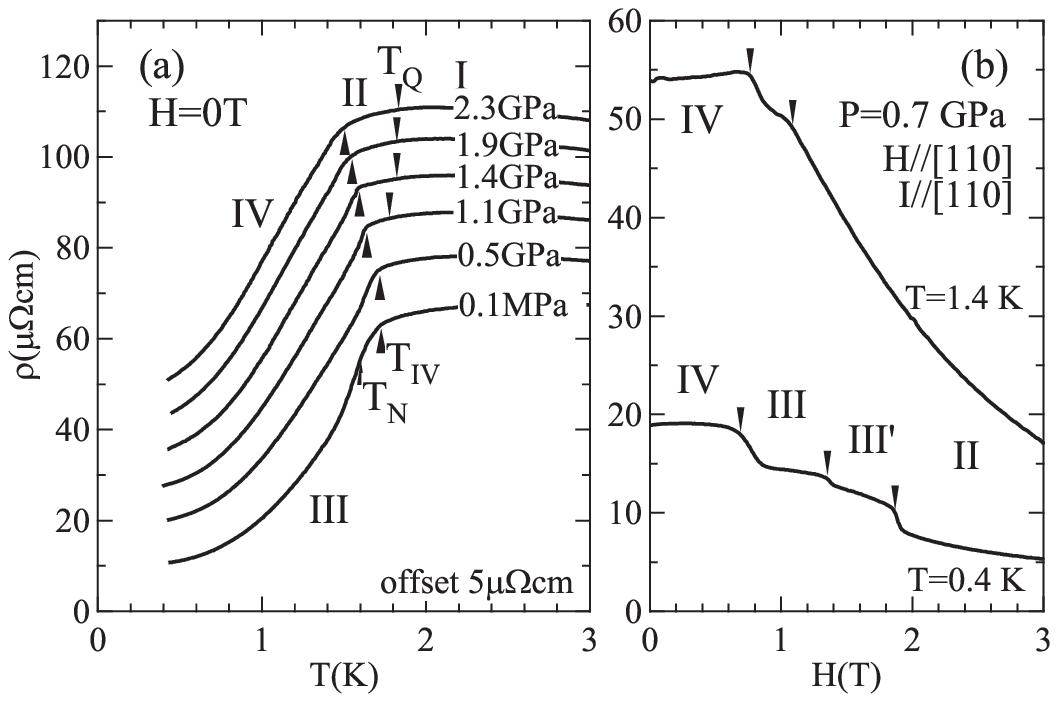}
\caption{Temperature dependence of the electrical resistivity of Ce$_{0.8}$La$_{0.2}$B$_{6}$ in the pressure range of up to $2.5$ GPa at $H = 0$ T. The origin of the horizontal axis is shifted for each set of data. (b) Magnetoresistance of Ce$_{0.8}$La$_{0.2}$B$_{6}$ under $P=0.7$ GPa for $H\parallel [110]$. }
\label{fig2}
\end{center}
\end{figure}

Figure \ref{fig1} shows the temperature ($T$) dependence of the specific heat of Ce$_{0.8}$La$_{0.2}$B$_{6}$ at $H=0$ and $0.6$ T for $H\parallel [001]$ at the ambient pressure. 
At $H=0$ T, two peaks are observed at $T\simeq 1.6$ and $1.7$ K. 
These correspond to the transition temperature from phase III to phase IV, $T_{\text{N}}$, and that from phase IV to phase I, $T_{\text{IV}}$, respectively. 
Hereafter, we define the transition temperature from phase IV to phase I and also that from phase IV to phase II as $T_{\text{IV}}$. 
At the ambient pressure, phase II does not exist at $H=0$ T. 
The temperature region of phase IV is $\sim 0.1$ K, which is narrower than that in Ce$_{0.75}$La$_{0.25}$B$_{6}$.\cite{ref6,ref7,ref8} 
Thus, Ce$_{0.8}$La$_{0.2}$B$_{6}$ is confirmed to be the sample in which phase IV only appears as a result of a higher decrease rate of $T_{\text{Q}}$ than that of $T_{\text{N}}$ owing to La substitution. 
At $H=0.6$ T, a sharp peak and a broad maximum are seen at $T\simeq 1.63$ and $\sim 2.0$ K, respectively. 
The former corresponds to $T_{\text{N}}$ and the latter to $T_{\text{Q}}$.

Figure \ref{fig2}(a) shows the $T$ dependence of the electrical resistivity ($\rho$) of Ce$_{0.8}$La$_{0.2}$B$_{6}$ under a pressure up to $2.3$ GPa at $H=0$ T. 
At the ambient pressure, two anomalies are observed at $T_{\text{N}}$ and $T_{\text{IV}}$. 
When pressure is applied, $T_{\text{N}}$ disappears rapidly and the ground state changes from phase III to phase IV. 
Although $T_{\text{IV}}$ is not changed up to $P=0.5$ GPa, it is suppressed when $T_{\text{Q}}$ appears above $T_{\text{IV}}$. 
While it is difficult to recognize $T_{\text{Q}}$ in the $\rho$ - $T$ curve in Fig. \ref{fig2}, it is confirmed by measurement in magnetic fields. 
Figure \ref{fig2}(b) shows the magnetoresistance under $P=0.7$ GPa for $H\parallel [110]$. 
The behavior of $\rho $ in phase IV of Ce$_{0.8}$La$_{0.2}$B$_{6}$ under pressure is the same as that of Ce$_{x}$La$_{1-x}$B$_{6}$ ($x \leq 0.7$) at the ambient pressure.\cite{ref19} 
This clearly shows that the ground state is phase IV without any influence from phase III, at least under $P=0.7$ GPa. 
The discontinuous changes of $\rho$ at the critical fields indicate the first-order phase transition at these critical fields.

\begin{figure}[t]
\begin{center}
\includegraphics[width=75mm,clip]{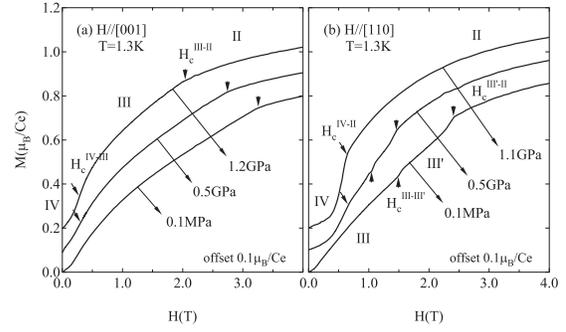}
\caption{Magnetization curves of Ce$_{0.8}$La$_{0.2}$B$_{6}$ at $T=1.3$ K for (a) $H\parallel [001]$ and (b) $H\parallel [110]$ under pressure. The origin of the horizontal axis is shifted for each set of data. }
\label{fig3}
\end{center}
\end{figure}

Figures \ref{fig3}(a) and \ref{fig3}(b) show the magnetization ($M$) curves of Ce$_{0.8}$La$_{0.2}$B$_{6}$ at $T=1.3$ K under pressure for $H\parallel [001]$ and $[110]$, respectively.
In the case of $H\parallel [001]$, at the ambient pressure, the domain redistribution from the three-domain state with $K_{xy}$, $K_{yz}$, and $K_{zx}$ to the single one with $K_{xy}$ occurs at $H\simeq 0.3$ T in phase III and the III-II phase transition is seen at $H_{\text{c}}^{\text{III-II}}\simeq 3.2$ T. 
Under $P=0.5$ GPa, $M$ shows a small concave $H$ dependence at around $H=0.2$ T, which is due to the appearance of phase IV. 
Under $P=1.2$ GPa, a more pronounced suppression of $M$ below $H\simeq 0.2$ T is seen and the increase in $M$ at $H\simeq 0.5$ T in phase III is also pronounced. 
$H_{\text{c}}^{\text{III-II}}$ is reduced from $\sim 3.2$ T at the ambient pressure to $\sim 2$ T under $P=1.2$ GPa. 
In the case of $H\parallel [110]$, at the ambient pressure, three anomalies are observed at $H\simeq 0.2$, $1.5$, and $2.4$ T. 
A small concave $H$ dependence of $M$ at around $H\simeq 0.2$ T is due to the domain redistribution from the three-domain state to the two-domain one with $K_{yz}$ and $K_{zx}$. 
The anomalies at $H\simeq 1.5$ and $2.4$ T correspond to the transition from phase III to phase III$^{\prime}$, $H_{\text{c}}^{\text{III-III}^{\prime}}$, and that from phase III$^{\prime}$ to phase II, $H_{\text{c}}^{\text{III}^{\prime}\text{-II}}$, respectively. 
The magnetic field region of phase III$^{\prime}$ at the ambient pressure expands greatly. 
Under $P=0.5$ GPa, $M$ at low magnetic fields is greatly suppressed. 
This indicates the appearance of phase IV in place of phase III. 
The IV-III, III-III$^{\prime}$, and III$^{\prime}$-II phase transitions occur at $H_{\text{c}}^{\text{IV-III}} \simeq 0.5$ T, $H_{\text{c}}^{\text{III-III}^{\prime}} \simeq 1$ T, and $H_{\text{c}}^{\text{III}^{\prime}\text{-II}} \simeq 1.4$ T, respectively. 
The critical fields $H_{\text{c}}^{\text{III-III}^{\prime}}$ and $H_{\text{c}}^{\text{III}^{\prime}\text{-II}}$ are greatly suppressed under a low pressure of $0.5$ GPa. 
Under $P=1.1$ GPa, the suppression of $M$ at low magnetic fields is much more pronounced than that under $P=0.5$ GPa. 
This clearly indicates that the ground state changes from phase III to phase IV.

\begin{figure}[t]
\begin{center}
\includegraphics[width=75mm,clip]{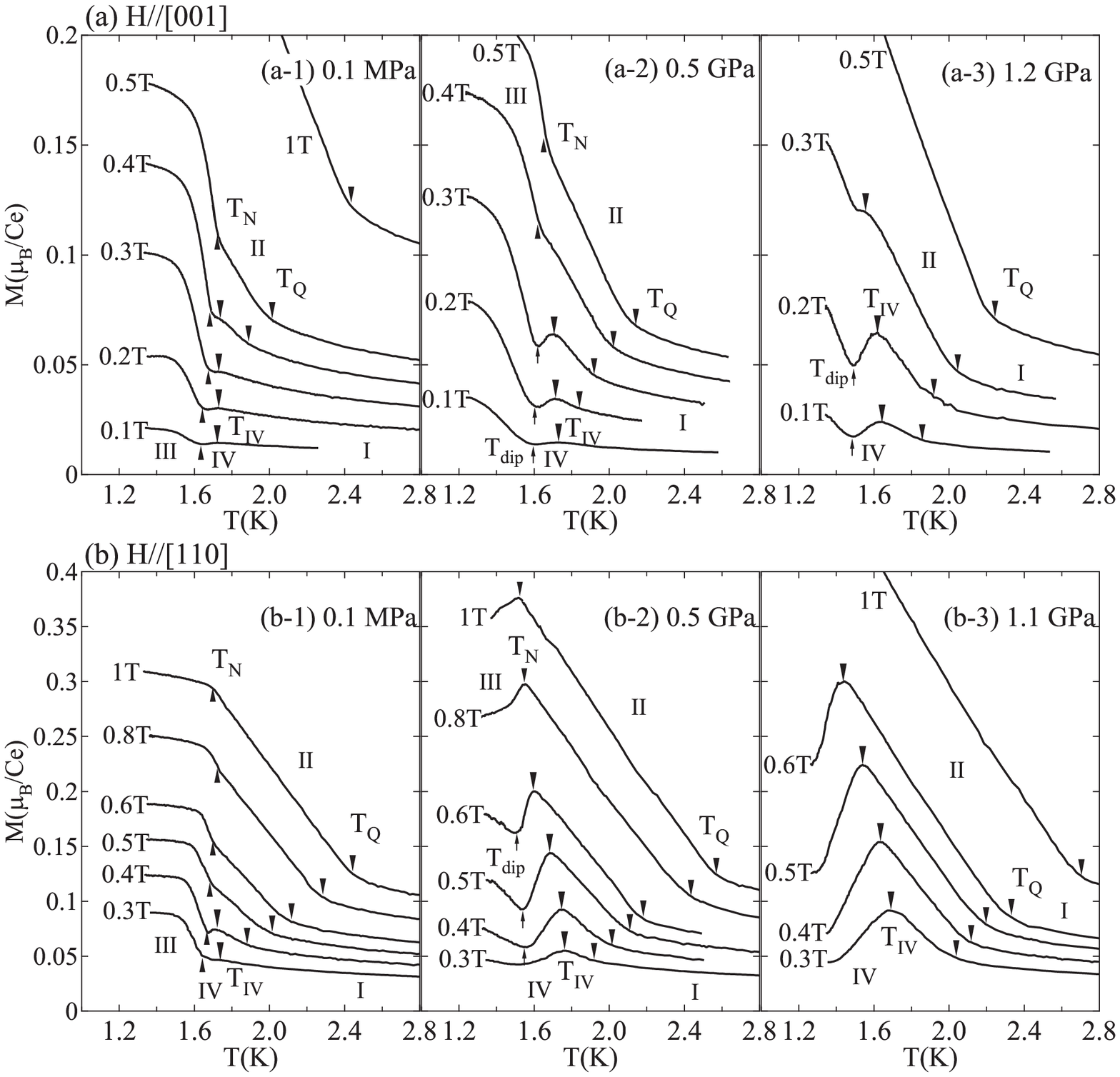}
\caption{Temperature dependence of magnetization of Ce$_{0.8}$La$_{0.2}$B$_{6}$ for (a) $H\parallel [001]$ and (b) $H\parallel [110]$ under pressure. Open arrows show the positions of $T_{\text{dip}}$.}
\label{fig4}
\includegraphics[width=73mm,clip]{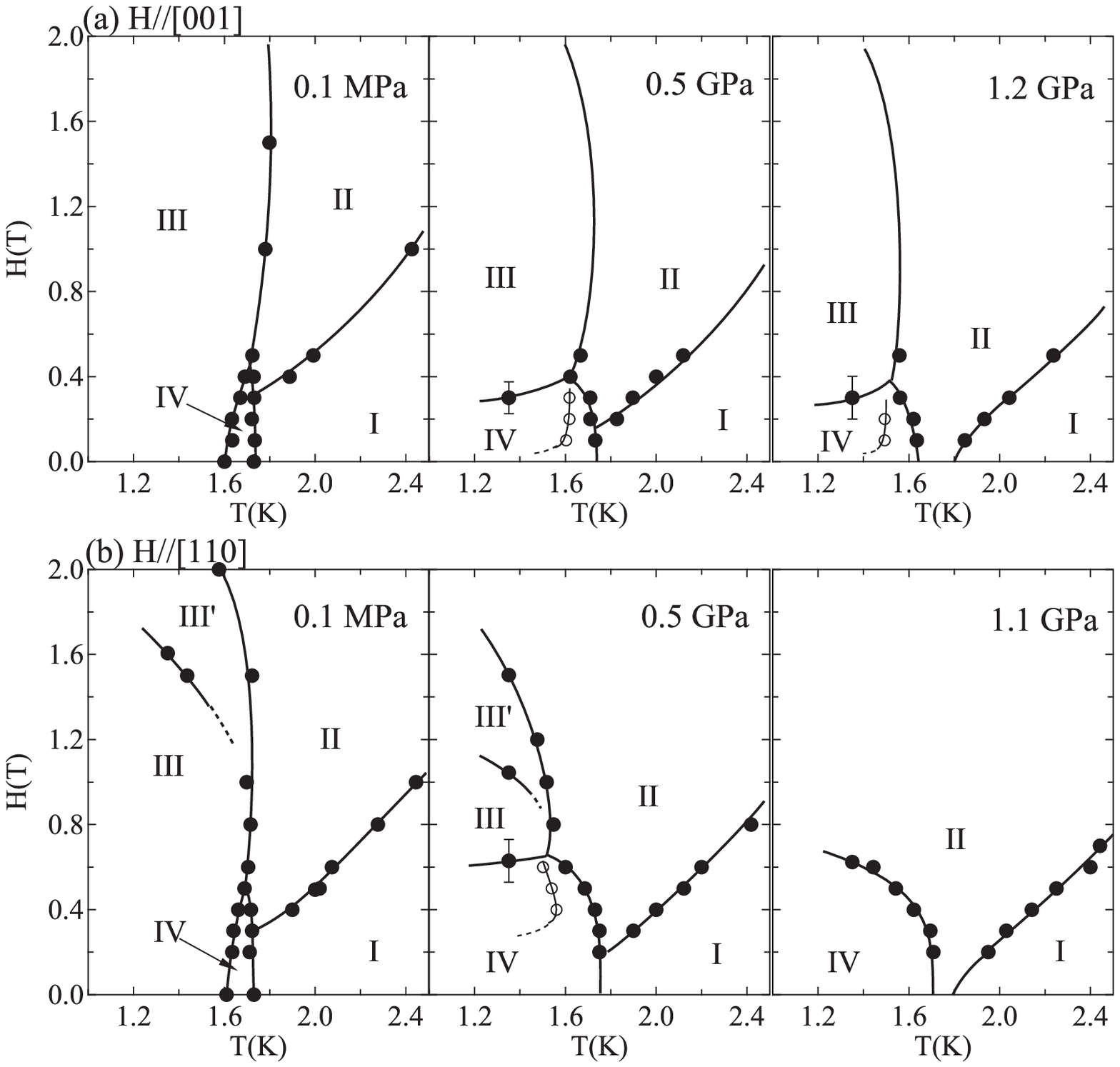}
\caption{Magnetic phase diagrams of Ce$_{0.8}$La$_{0.2}$B$_{6}$ for (a) $H\parallel [001]$ and (b) $H\parallel [110]$ under pressure. Open circles indicate results due to the dip in the $M$-$T$ curve. See the text for details. }
\label{fig5}
\end{center}
\end{figure}

Figures \ref{fig4}(a-1) - \ref{fig4}(a-3) show the $T$ dependence of $M$ for $H\parallel [001]$ in the pressure range of up to $P=1.2$ GPa. 
In the present case, a peak in the $M$-$T$ curve is always observed at $T_{\text{IV}}$. 
This is different from the results obtained under the uniaxial pressure $P\parallel [001]$ for $H\parallel [001]$, where $M$ increases with decreasing temperature below $T_{\text{IV}}$.\cite{ref16} 
Figure \ref{fig4}(a-1) shows the $M$-$T$ curves at the ambient pressure for $H\parallel [001]$. 
At $H=0.1$ T, $M$ shows a small peak at $T_{\text{IV}}$ and a dip at $T_{\text{N}}\simeq 1.6$ K, which correspond to the two peaks observed in the specific heat in Fig. \ref{fig1}. 
Below $T\simeq 1.5$ K, $M$ is almost independent of temperature. 
The existence of $T_{\text{IV}}$ is recognized up to $H=0.4$ T, and at $H=0.4$ T, $T_{\text{Q}}$ is recognized at $\sim 1.9$ K. 
Figure \ref{fig4}(a-2) shows the $M$-$T$ curves under $P=0.5$ GPa. 
At $H=0.1$ T, $M$ shows a small peak at $T_{\text{IV}}\simeq 1.75$ K and a dip at $T_{\text{dip}}\simeq 1.6$ K. 
At $H=0.2$ and $0.3$ T, $T_{\text{Q}}$ appears in addition to $T_{\text{IV}}$ and $T_{\text{dip}}$ at higher temperatures. 
Above $H=0.4$ T, there exist only $T_{\text{Q}}$ and $T_{\text{N}}$. 
Figure \ref{fig4}(a-3) shows the $M$-$T$ curves under $P=1.2$ GPa. 
$T_{\text{Q}}$ is recognized already at $H=0.1$ T, where $M$ shows a large increase below $T_{\text{Q}}\simeq 1.85$ K and a clear peak at $T_{\text{IV}} \simeq 1.65$ K. 
At $H=0.2$ T, the peak at $T_{\text{IV}}$ becomes more pronounced. 
At $H=0.3$ T, the peak is negligible at $T_{\text{IV}}\simeq 1.55$ K and $M$ shows a large increase below $T\simeq 1.5$ K, which is due to the entrance into phase III.

Figures \ref{fig4}(b-1) - \ref{fig4}(b-3) show the $T$ dependence of $M$ for $H\parallel [110]$ in the pressure range of up to $P=1.1$ GPa. 
Figure \ref{fig4}(b-1) shows the $M$-$T$ curves at the ambient pressure. 
At $H=0.3$ T, $M$ shows a small peak at $T_{\text{IV}} \simeq 1.75$ K and an increase below $T_{\text{N}} \simeq 1.65$ K. 
Below $T \simeq 1.5$ K, $M$ is almost independent of temperature. 
At $H=0.4$ T, $T_{\text{Q}}$ is recognized at $\sim 1.85$ K, and a peak exists at $T_{\text{IV}}\simeq 1.73$ K. 
Above $H=0.5$ T, only two anomalies are seen at $T_{\text{Q}}$ and $T_{\text{N}}$. 
Figure \ref{fig4}(b-2) shows the $M$-$T$ curves under $P=0.5$ GPa. 
The results are similar to those obtained under the uniaxial pressure $P\parallel [001]$ for $H\parallel [110]$.\cite{ref16} 
At $H=0.3$ T, $M$ shows an increase below $T_{\text{Q}}\simeq 1.9$ K and a peak at $T_{\text{IV}}\simeq 1.75$ K. 
Below $T_{\text{IV}}$, no clear anomaly is seen down to $T\simeq 1.3$ K. 
At $H=0.4$ T, the increase in $M$ below $T_{\text{Q}}$ is much more clear than that at $H=0.3$ T. 
$M$ shows a clear peak at $T_{\text{IV}}\simeq 1.75$ K and a clear dip at $T_{\text{dip}}\simeq 1.6$ K, which is different from the result obtained at $H=0.3$ T. 
Above $H=0.8$ T, after showing a large increase below $T_{\text{Q}}$, $M$ shows a peak at $T_{\text{N}}$. 
Here, we note the difference between $T_{\text{N}}$ and $T_{\text{dip}}$. 
While the increases in $M$ below both $T_{\text{N}}$ and $T_{\text{dip}}$ are of the same type, these two temperatures are different. 
Below $T_{\text{N}}$, the pure phase III is realized. 
However, the ordered phase below $T_{\text{dip}}$ exibits the characteristics of both phase IV and phase III, which is seen at a magnitude of $M$ below $T_{\text{dip}}$ under $P=0.5$ GPa that is smaller than that below $T_{\text{N}}$ at the ambient pressure, as is seen in Fig. \ref{fig3}(b).
Figure \ref{fig4}(b-3) shows the $M$-$T$ curves under $P=1.1$ GPa. 
In the present case, $M$ shows a large increase below $T_{\text{Q}}\simeq 2.1$ K at $H=0.3$ T and a peak at $T_{\text{IV}}\simeq 1.7$ K. 
The peak of $M$ at $T_{\text{IV}}$ is clearly recognized up to $H=0.6$ T.

Figures \ref{fig5}(a) and \ref{fig5}(b) show the magnetic phase diagrams of Ce$_{0.8}$La$_{0.2}$B$_{6}$ for $H\parallel [001]$ and $H\parallel [110]$, respectively. 
The results obtained at the ambient pressure are the same as those reported previously.\cite{ref9} 
Phase III is rapidly suppressed by applying only a small pressure of $P=0.5$ GPa and the ground state becomes phase IV. 
These are also observed in the electrical resistivity under pressure in Fig. \ref{fig2}. 
With increasing pressure, $T_{\text{Q}}$, which exists in a finite magnetic field at the ambient pressure, is reduced to lower magnetic fields. 
Above $P\simeq 1.1$ GPa, $T_{\text{Q}}$ appears at $H=0$ T at the higher temperature side of $T_{\text{IV}}$. 
Thus, the IV-I phase transition disappears and the successive phase transitions from phase IV to phase II and from phase II to phase I appear at $H=0$ T above $P\simeq 1.1$ GPa. 
The magnetic phase diagram under pressure shows a large anisotropy depending on the applied field direction. 
For $H\parallel [001]$, the suppression of phase III under pressure is much less than that for $H\parallel [110]$. 
The reduction in $T_{\text{N}}$ is also small in magnetic fields. 
As a result, the temperature region where the IV-II direct phase transition exists is narrow even under $P=1.2$ GPa for $H\parallel [001]$. 
$H_{\text{c}}^{\text{IV-III}}$ is also small under pressure for $H\parallel [001]$. 
In the case of $H\parallel [110]$, phase III in finite magnetic fields is rapidly suppressed. 
$H_{\text{c}}^{\text{IV-II}}$ is rapidly reduced, and $T_{\text{N}}$ is also reduced to lower temperatures. 
The temperature region where the IV-II direct phase transition exists expands with increasing pressure. 
$H_{\text{c}}^{\text{IV-II}}$ is about $\sim 0.6$ T at $T=1.3$ K under $P=1.1$ GPa.


Now, we discuss the present results. 
First, we briefly discuss the rapid suppression of phase III under pressure. 
In our previous paper\cite{ref20}, we reported that the crystal volume is larger in phase III than in phase II at $H=0$ T. 
The expansion of the crystal volume in phase III may be ascribed to the destruction of the Kondo effect, considering that the Kondo state gains energy owing to a $c$-$f$ mixing effect, which leads to the shrinkage of the crystal volume. 
In the present study, it is found that $\mathrm{d} T_{\text{IV}} / \mathrm{d} P$ is small, at least for $P \leq 0.5$ GPa where the direct transition from phase IV to phase I occurs. 
The rapid suppression of phase III caused by applying a low pressure means that phase III with a larger crystal volume shows difficulty in continuously existing inside phase IV, at least at $H=0$ T under pressure.

Next, we discuss the IV-II direct phase transition. 
We found that the successive transitions from phase IV to phase II and from phase II to phase I appear at $H=0$ T under pressure above $\sim 1.1$ GPa. 
This is because phase II, which exists above $H \simeq 0.4$ T at the ambient pressure, is stabilized under pressure, and finally, above $P \simeq 1.1$ GPa, phase II appears at $H=0$ T at the higher temperature side of phase IV. 
Thus, the situation in which phase IV exists stably inside phase II is realized. 
On the other hand, Kondo \textit{et al}. showed that the $O_{xy}$-type FQ ordering simultaneously induced by the $\Gamma_{\text{5u}}$-type AFO ordering is easily suppressed by introducing the $O_{xy}$-type AFQ interaction that should exist in a Ce$_{x}$La$_{1-x}$B$_{6}$ system within the mean-field calculation framework where the four interactions of the $\Gamma_{\text{5u}}$-type AFO, $O_{xy}$-type AFQ, $T_{xyz}$-type AFO, and AF exchange are taken into account.\cite{ref14} 
The suppression is due to the fact that the $O_{xy}$-type FQ ordering simultaneously induced in the $\Gamma_{\text{5u}}$-type AFO ordering competes with the $O_{xy}$-type AFQ interaction, and it is considered that when $T_{\text{Q}}$ is higher than the $\Gamma_{\text{5u}}$-type AFO ordering temperature, the $\Gamma_{\text{5u}}$-type AFO ordering disappears. 
However, the present experimental results show that phase IV exists stably inside phase II. 
Thus, the present results seem to be difficult to reproduce by introducing the above four interactions within the mean-field calculation framework.

Finally, we discuss the effect of phase III on phase IV at low magnetic fields under pressure. 
First, we note the difference between the uniaxial pressure effects studied by Sakakibara \textit{et al}.\cite{ref16} and the hydrostatic pressure effects presented here. 
In the case of the uniaxial pressure $P\parallel [001]$, a $K_{xy}$ single-domain state can be introduced into phase IV because the $K_{xy}$ domain is expected to be tetragonal and the $z$-axis is shorter than the $x$- and $y$-axes.\cite{ref20} 
If the $K_{xy}$ domain mixes with phase IV owing to the first-order transition between phase IV and phase III as discussed by Sakakibara \textit{et al}.\cite{ref16}, it seems possible to explain the experimental results under the uniaxial pressure $P\parallel [001]$ both for $H\parallel [001]$ and $[110]$. 
On the other hand, under a hydrostatic pressure, as phase III disappears rapidly upon applying a low pressure, the ground state changes from phase III to phase IV and phase III may not coexist with phase IV, at least at $H=0$ T. 
However, the magnetization curves at low magnetic fields under the highest pressure studied here seem to indicate that phase III is induced for $H\parallel [001]$ but not for $H\parallel [110]$. 
$T_{\text{dip}}$ in the $M$-$T$ curve under pressure in Fig. \ref{fig4} seems to divide phase IV in finite magnetic fields into two regions. 
One is pure phase IV between $T_{\text{dip}}$ and $T_{\text{IV}}$ and the other is the mixed phases of III and IV below $T_{\text{dip}}$. 
Here, we point out that the reason why phase III is introduced into phase IV under pressure at low magnetic fields is not because the magnitude of $M$ in phase III is larger than that in phase IV. 
This is verified by the fact that although $M$ in phase II is larger than that in phase IV, phase II is not introduced into phase IV. 
This is clearly seen in the $M$-$H$ curve for $H\parallel [110]$ under $P=1.1$ GPa where there is a clear transition from phase IV with a small $M$ to phase II with a large $M$. 
Here, we point out that AFM ordering exists in phase III but not in phase II. 
Considering that phase III is introduced into phase IV for $H\parallel [001]$ but not for $H\parallel [110]$, whether phase III can be introduced into phase IV or not depends on the different nature of phase III governed by the applied magnetic field direction. 
In the $K_{xy}$ domain for $H\parallel [001]$, the AF magnetic moments lie in the $xy$-plane, and in the $K_{yz}$ or $K_{zx}$ domain for $H\parallel [110]$, they lie along the $z$-direction. 
This difference seems to be associated with the origin of the introduction or nonintroduction of phase III into phase IV being dependent on the applied magnetic field direction. 
These results may provide an important clue to understand the relationship between phase IV and phase III.


In conclusion, we have studied the pressure effect of the electrical resistivity and magnetization of Ce$_{0.8}$La$_{0.2}$B$_{6}$. 
By lowering $T_{\text{Q}}$ to zero magnetic field by applying pressure, the situation in which phase IV exists inside phase II could be realized. 
The obtained results seem to be difficult to reproduce by taking the four interactions of $\Gamma_{\text{5u}}$-type AFO, $O_{xy}$-type AFQ, $T_{xyz}$-type AFO, and AF exchange into account within the mean-field calculation framework. 
We also investigated the effect of phase III on phase IV at low magnetic fields under pressure and found that phase III is easily introduced into phase IV for $H\parallel [001]$ but not for $H\parallel [110]$.


\newpage

\end{document}